\documentclass[aps,prd,superscriptaddress,preprint,tightenlines,nofootinbib]{revtex4}

\usepackage{graphicx}
\usepackage{dcolumn}
\usepackage{bm}

\newcommand{\bpm}{{\cal B}_{+-}}
\newcommand{\ccbar}{c \bar c}
\newcommand{\cm}{\,\mbox{cm}}
\newcommand{\gev}{\,\mbox{GeV}}
\newcommand{\mev}{\,\mbox{MeV}}
\newcommand{\ntrk}{N_{\mathrm{trk}}}
\newcommand{\chisqv}{\chi^2_{\mathrm{V}}/{\mathrm {d.o.f.}}}
\newcommand{\chisqm}{\chi^2_{\mathrm{M}}/{\mathrm {d.o.f.}}}
\newcommand{\invpb}{\,\mbox{pb}^{-1}}
\newcommand{\dilep}{\ell^+\ell^-}
\newcommand{\diel}{e^+e^-}
\newcommand{\dimu}{\mu^+\mu^-}
\newcommand{\jpsi}{J/\psi}
\newcommand{\piz}{\pi^0}
\newcommand{\dipi}{\pi^+\pi^-}
\newcommand{\dipiz}{\piz\piz}
\newcommand{\pp}{\psi(2S)}
\newcommand{\chicJ}{\chi_{cJ}}
\newcommand{\egammalow}{E_{\gamma\mathrm{\mbox{-}low}}}
\newcommand{\evis}{E_{\mathrm{vis}}}
\newcommand{\ecm}{E_{\mathrm{cm}}}
\newcommand{\ebeam}{E_{\mathrm{beam}}}
\newcommand{\jany}{\mathrm{any} + \jpsi}

\newcommand{\ggnr}{(\gamma\gamma)_{\mathrm{nr}}}
\newcommand{\symbols}{Symbols are as in Fig.~1. The dotted line represents the simulated signal. }

\begin{document}

\title{Branching Fractions for Transitions of 
\boldmath $\psi(2S)$ to $\jpsi$}

\preprint{CLNS 08/2025}
\preprint{CLEO 08-08}

\author{H.~Mendez}
\affiliation{University of Puerto Rico, Mayaguez, Puerto Rico 00681}
\author{J.~Y.~Ge}
\author{D.~H.~Miller}
\author{I.~P.~J.~Shipsey}
\author{B.~Xin}
\affiliation{Purdue University, West Lafayette, Indiana 47907, USA}
\author{G.~S.~Adams}
\author{M.~Anderson}
\author{J.~P.~Cummings}
\author{I.~Danko}
\author{D.~Hu}
\author{B.~Moziak}
\author{J.~Napolitano}
\affiliation{Rensselaer Polytechnic Institute, Troy, New York 12180, USA}
% ...
\collaboration{CLEO Collaboration}
\noaffiliation

\date{April 28, 2008}

\begin{abstract} 
We report determination of branching fractions for the
decays $\psi(2S) \to h + \jpsi$,
where $h=\mathrm{any}$, $\dipi$, $\dipiz$, $\eta$, $\piz$,
and $\gamma\gamma$ through $\chi_{c0,1,2}$.
These measurements use 27M $\psi(2S)$ decays produced in $\diel$ 
collision data collected with the 
CLEO detector. 
The resulting branching fractions and 
ratios thereof improve upon previously achieved
precision in all cases, and in combination with other
measurements permit determination of ${\cal B}(\chicJ \to \gamma \jpsi)$
and ${\cal B} (\psi(2S) \to $ light hadrons$)$.
\end{abstract}

\pacs{13.25.Gv,13.20.Gd}
\maketitle

The study of
charmonium has entered an era in which
many of the broad features are finally
known, increasingly focusing attention upon the details
of both production and decay~\cite{voloshin}. The decays of
$\psi(2S)$ in particular have become very well studied.
About 4/5 of all $\psi(2S)$ decays are through de-excitation,
mostly by hadronic transition to the $\jpsi$, but also through
radiative decay to the $\chicJ$ states. 
Measurements of rates for such processes
enable meaningful comparison
with theory and extrapolation to similar
mechanisms in the $\Upsilon$ system. 
The listed transitions can be used to isolate and
study lower-lying $\ccbar$ states. 
For example, the decay $\psi(2S) \to \dipi\jpsi$ is
useful because this is the most common and
the most accessible experimentally; 
inclusive and exclusive $\jpsi$
decays are often identified by tagging the recoil dipion. 
The corresponding $\psi(2S)$ decay rate is therefore
a target for continuing precision improvements.
Similarly, $\chicJ$ mesons can often be tagged for study
by the transition photons.
The {\em inclusive} rate for transitions to the $\jpsi$ is
also a crucial input to predictions
for $\ccbar$ annihilation rates because it
limits the remainder. Hence refining the
rates for $\psi(2S)$-to-$\jpsi$ transitions remains
an experimental priority. The first comprehensive look
at all transitions simultaneously by a single
experiment was reported by CLEO~\cite{xjpsi_3m}
in 2005; absolute measurements were limited
by a 3\% uncertainty in the number of $\psi(2S)$ produced,
and many ratios of rates were limited
by statistics. Further investigations with
a larger dataset and improved systematic uncertainties 
are certainly warranted.

This article describes a measurement of ratios
of branching fractions $\psi(2S) \to h + J/\psi$, where
$h= \dipi$, $\dipiz$, $\eta(\to \gamma\gamma, \dipi\piz)$, 
$\piz$, $\gamma\gamma$ through $\gamma \chicJ$ (for which process
we will also use the expression $\gamma (\gamma\jpsi)_{\chicJ}$),
and the inclusive
branching fraction ${\cal B}(\psi(2S) \to \jany)$, in which only
the $\jpsi$ is identified.
We separately determine
$\bpm =  {\cal B}(\psi(2S) \to \dipi\jpsi)$ with the $\jpsi$
decaying inclusively, which we denote by 
$\jpsi \to X$.

We use $e^+e^-$~collision data at and below the $\pp$~resonance,
$\ecm = 3.686\gev$ ($\int {\cal L} dt = 53.8\invpb$, corresponding to 
27M~$\psi(2S)$ decays) and
$\ecm = 3.670\gev$ (``continuum'' data, $\int {\cal L}dt = 20.6\invpb$).
Events were acquired at the Cornell Electron Storage Ring (CESR)~\cite{cesr} 
with the CLEO detector~\cite{cleo}, mostly in the CLEO-c 
configuration (95\%), with the balance from CLEO~III. 
The detector features a solid angle coverage of $93\%$ for
charged and neutral particles. 
The charged particle tracking system operates in a 1.0~T~magnetic field
along the beam axis and achieves a momentum resolution of
$\sim 0.6\%$ at momenta of $1\gev/c$. The CsI crystal
calorimeter attains
photon energy resolutions of $2.2\%$ for $E_\gamma = 1\gev$
and $5\%$ at $100\mev$.

For the measurement of branching fractions relative to
one another, the $\jpsi$ is identified through its decay
to $\dimu$ or $\diel$. 
We require $|\cos\theta_{\mathrm{trk}}|< 0.83$ for both lepton tracks,
where the polar angle $\theta$ is measured with respect to the 
positron direction of incidence. 
The ratios of calorimeter shower energy to track momentum, $E/p$, 
for the lepton candidates, taken to be the two 
tracks of highest momentum in the event, 
must be larger than $0.85$ for one electron
and above $0.5$ for the other, or in the case of muons
smaller than $0.25$ for one and below $0.5$ for the other. 
In order to salvage lepton pairs that
have radiated photons and would hence lose too much 
energy to remain identifiable as a $\jpsi$,
we add bremsstrahlung photon candidates
found within a cone of 100$\,$mrad to the 
three-vector of each lepton track at the interaction point (IP).
The $\jpsi$ candidate is retained only if constrained fits
to the two tracks and bremsstrahlung candidates
to a common vertex and to the mass of the $\jpsi$ 
fulfil $\chisqv < 20$ and $\chisqm < 20$, respectively.
The $\chisqm$ restriction corresponds roughly to demanding
that the dilepton mass lie between 3.03 and 3.16~GeV. 

For $\psi(2S) \to \jany$, cosmic ray background is rejected 
based on the distance of the track impact
parameters to the event interaction point ($<2\,$mm), and
on the $\jpsi$~momentum ($p_{\jpsi}>50\mev/c$). To suppress
background from continuum reactions 
we require $p_{\jpsi} < 570\mev$.
Radiative lepton pair production 
and radiative returns to the $\jpsi$ ($\diel \to \gamma \jpsi$)
are suppressed by requiring $|\cos\theta_{J/\psi}|<0.98$ and
for the dielectron mode by demanding $\cos\theta_{e^+}< 0.5$. 
Decays of $\psi(2S)$ to final states not involving a $J/\psi$ can
contaminate the $\jany$ mode if two oppositely charged
particles satisfy the lepton identification and
kinematic criteria. Monte Carlo (MC) studies indicate such 
backgrounds are very small, leading to no background subtraction and
an assignment of a systematic uncertainty in rate of 0.2\% ($\dimu$)
and 0.1\% ($\diel$).

For $\dipi\jpsi$ [$\dipiz\jpsi$] we demand the presence 
of two oppositely charged tracks [two $\piz$ candidates].
In both cases, the mass
recoiling against the dipion is required to be 3.05-3.15~GeV.
To suppress background from radiative
Bhabha and dimuon pairs, where the photon converts and the
resulting tracks are associated with the $\dipi$, 
we require $m(\dipi) > 0.35\mev$.

Any photon candidate reconstructed in 
any of the exclusive modes must lie in the central region of the
calorimeter ($|\cos \theta_\gamma| < 0.75$) and be isolated
(not aligned with the initial momentum of a track
within 100~mrad and not closer than 30~cm to a shower that
is matched to a track). We impose mode-dependent requirements on
the photon energies, $E_\gamma$. 

All $\piz$ candidates are formed using two photon showers,
each of at least $30\mev$ in energy, that together have an 
invariant mass of $100$-$160\mev$. 
For $\dipiz\jpsi$ and $(\dipi\piz)_\eta\jpsi$,
but not for $\piz\jpsi$,
we constrain their four-momentum to the $\piz$ mass.
For $\piz\jpsi$, the softer of the two photons 
cannot lie within $E_\gamma = 100$-$200~\mev$ in
order to suppress background from $\gamma \chicJ$ decays.

To select $(\dipi\piz)_{\eta}$ events, we require two
oppositely-charged tracks and a $\piz$ candidate, 
which together have an invariant mass of $535$-$560\mev$.
For $(\gamma\gamma)_\eta \jpsi$, we require that both
photons have $E_\gamma > 200\mev$ and $m(\gamma\gamma) = 500$-$600\mev$.
Leakage of the high-rate process
$\dipiz\jpsi$ into the other modes with only two 
photons and no tracks is suppressed by
requiring that for $\piz\jpsi$ and $\gamma\chicJ$ the third-largest
shower energy in the event does not exceed $50\mev$. 
Both $\eta$ decay modes are combined into a common $\eta$ measurement
using their branching fractions~\cite{etabr}.

For $\piz\jpsi$ [$\eta\jpsi$], we only keep events where 
$\jpsi$ candidates have a momentum within 490-570~MeV/$c$ 
[170-230]~MeV/$c$.

For $\gamma(\gamma\jpsi)_{\chicJ}$,
we tally yields, cross-feeds, and backgrounds separately
for the following windows applicable to $\egammalow$, the
lower photon energy of the two:
90-145~MeV ($\chi_{c2}$),
145-200 ($\chi_{c1}$),
200-245 (between $\chi_{c1}$ and $\chi_{c0}$),
and
245-290 ($\chi_{c0}$).
For the last two categories, we subject the event to a final 
kinematic fit constraining the $\jpsi$ and the two
photons to the $\psi(2S)$ four-momentum. 
The $\jpsi$
momentum must lie in the range $250$-$500\mev$ to eliminate
$\piz\jpsi$ and $(\gamma\gamma)_\eta \jpsi$ cross-feed,
and for $J=1,2$, 
the invariant mass recoiling against the two photons
must lie near the $\jpsi$ mass, 3.05-3.13~GeV.

\begin{figure}[t]
\includegraphics*[width=6.5in]{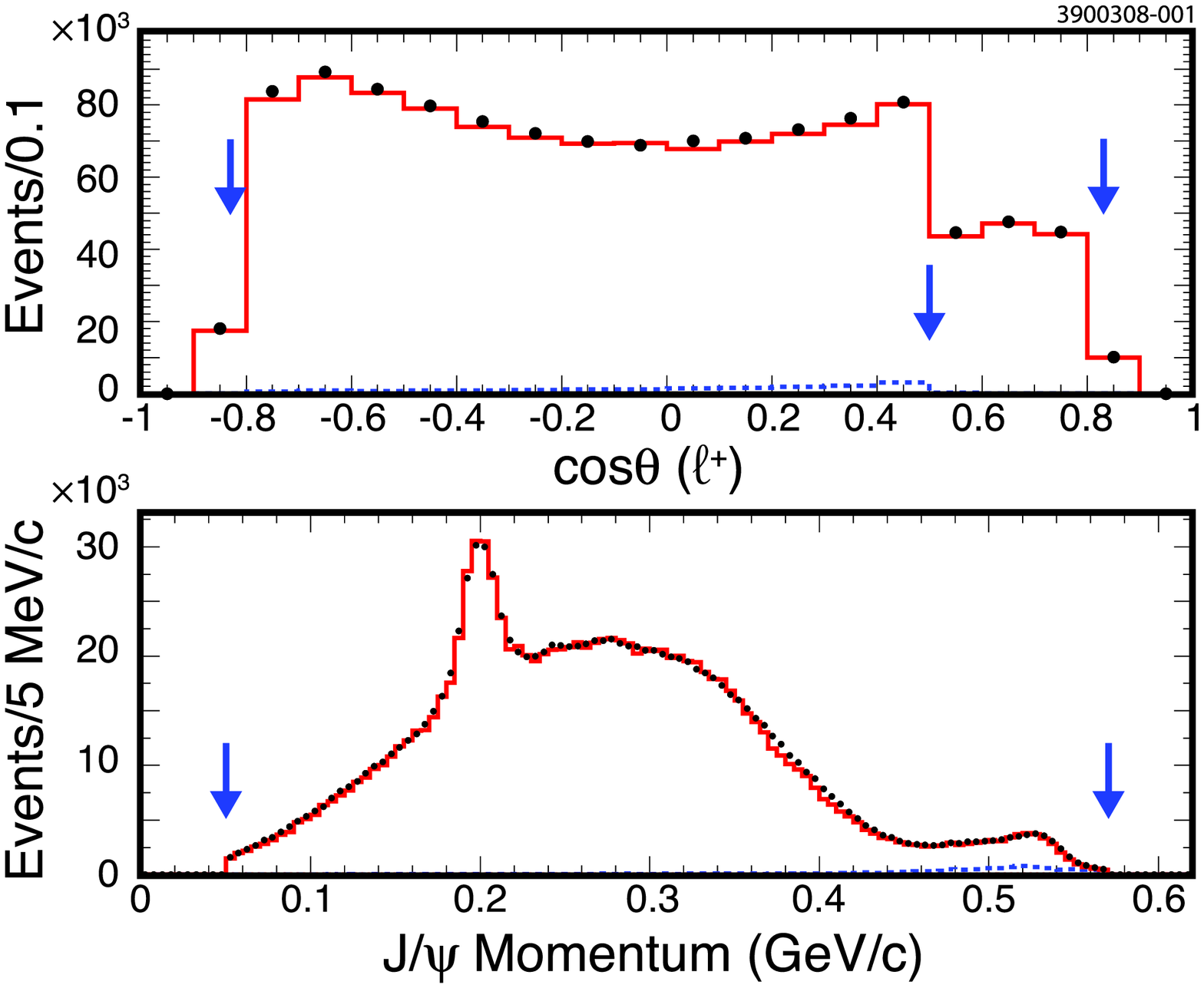}
\caption{Distributions relevant to $\jany$. 
Top: polar angle of the positive lepton. 
For $\diel$ only,
we demand $\cos\theta_{e^+} < 0.5$ 
to suppress Bhabha events with initial/final state radiation
or bremsstrahlung in detector material.
Bottom: $\jpsi$ momentum. 
Solid circles show the on-$\psi(2S)$ data,
dashed histogram the continuum
data (scaled by luminosity and 1/$s$)
taken at $\ecm =$3.67~GeV,
the solid histogram represents the sum of
all MC exclusive channels (scaled
to match the data in signal modes). 
Arrows appear
at nominal selection values. 
\label{fig:jany} }
\end{figure}

\begin{figure}[t]
\includegraphics*[width=6.5in]{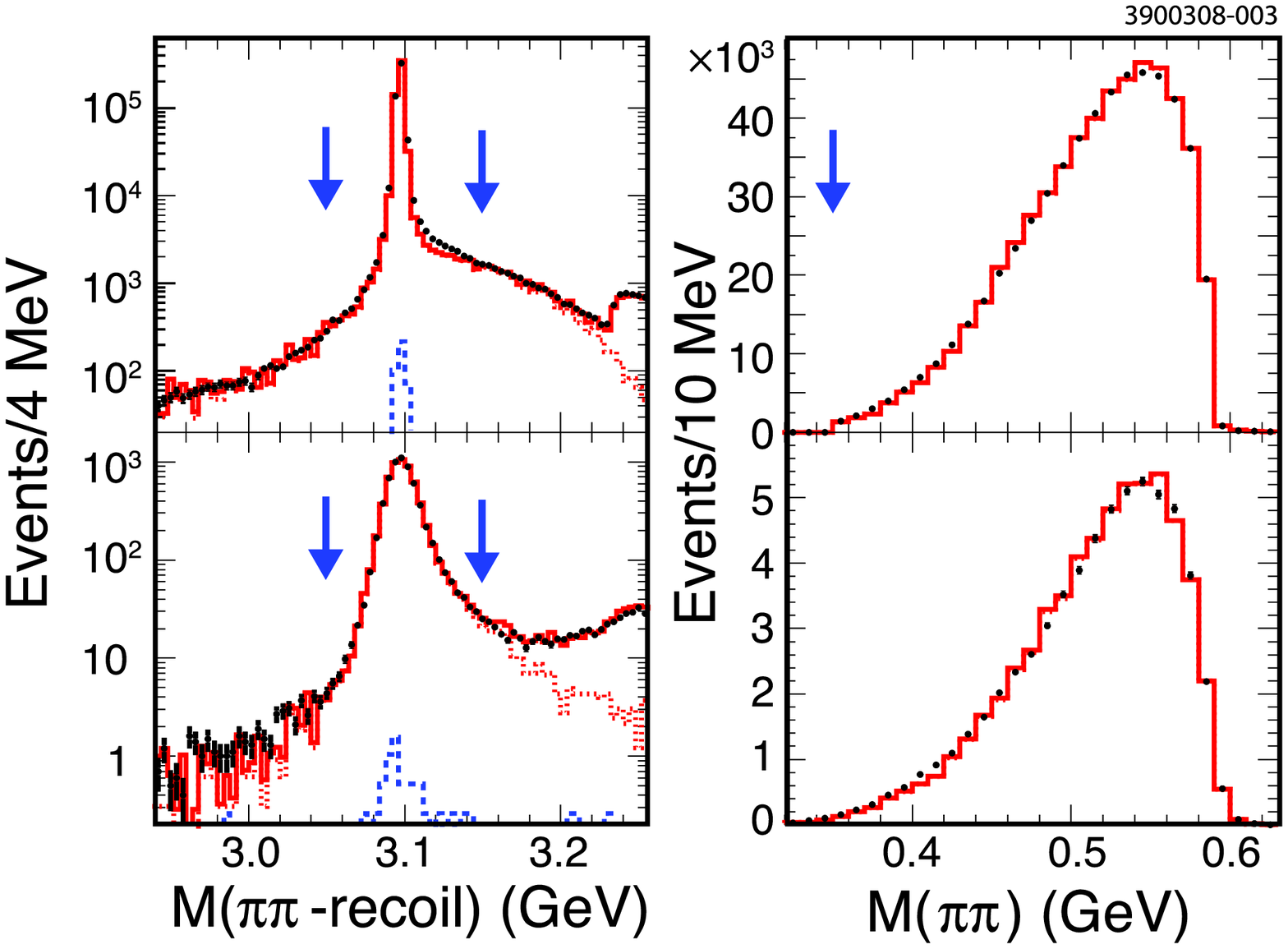}
\caption{Plots relevant to the decay
$\pp\to\dipi\jpsi$ (top) and
$\pp\to\dipiz\jpsi$ (bottom).
The left plots show the dipion
recoil mass spectrum and the
right plots the dipion mass spectrum.
The $J/\psi$ candidates in the continuum sample arise from
the tail of the $\psi(2S)$.
\symbols
\label{fig:pipij} }
\end{figure}

\begin{figure}[t]
\includegraphics*[width=6.5in]{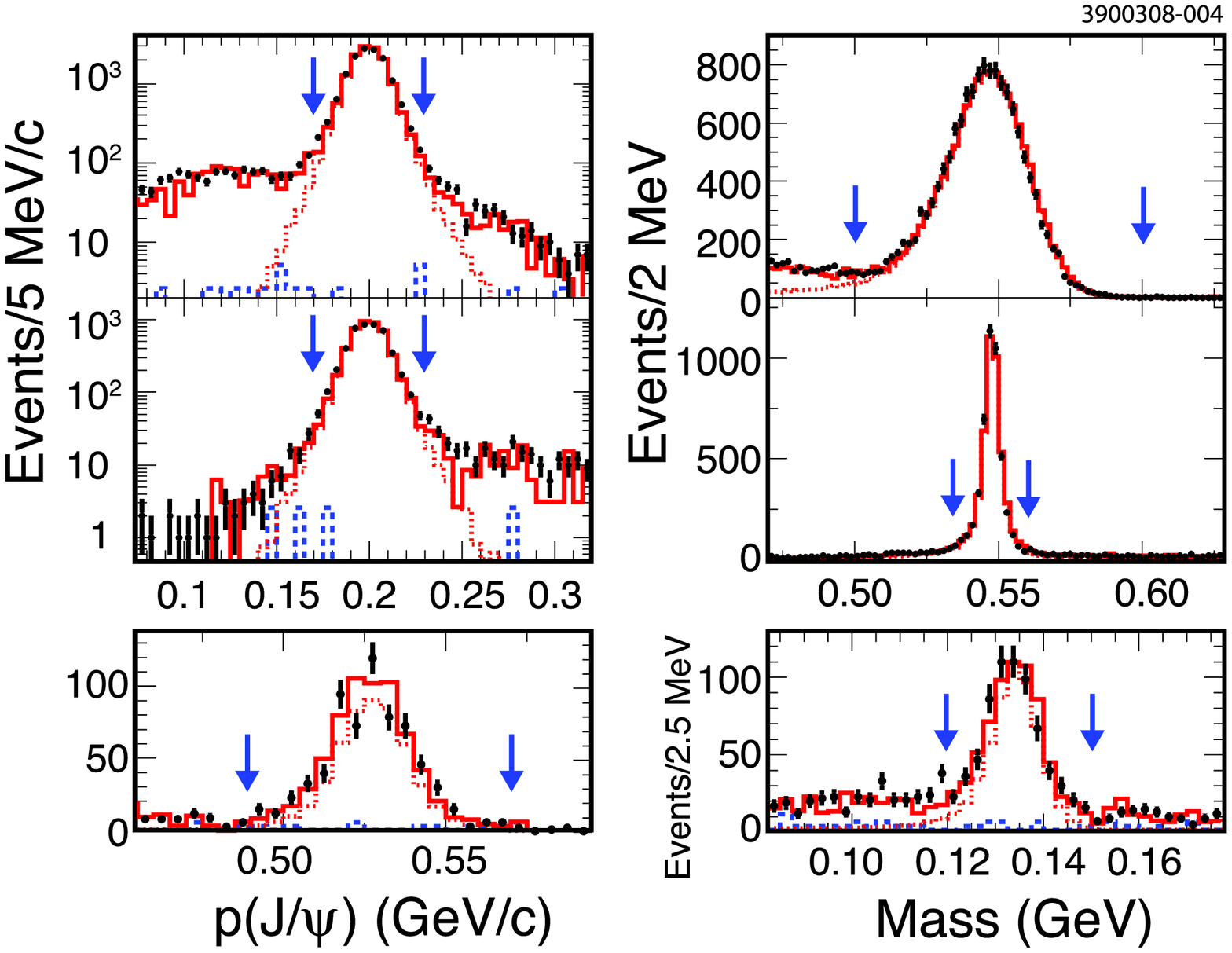}
\caption{
$\pp\to\eta\jpsi$, $\eta\to\gamma\gamma$ (top)
and $\eta\to\dipi\piz$ (middle), and
$\pp \to \piz\jpsi$ (bottom):
The $\jpsi$ momentum (left) and invariant mass of
the decay products (right).
\symbols\ 
\label{fig:etaj_and_pi0j} }
\end{figure}

After these requirements, the event samples are clean. This is
demonstrated for $\jany$, $\dipi\jpsi$ and $\dipiz\jpsi$,
as well as $\eta\jpsi$ and $\piz\jpsi$ in Figs.~\ref{fig:jany},
\ref{fig:pipij}, and \ref{fig:etaj_and_pi0j}.
The remaining background is readily estimated and subtracted.
The most important backgrounds
come from cross-feed among our signal modes, for which we account 
using the measured yield from our sample as background
normalization. We also calculate background from $\eta\jpsi$
with $\eta$ decaying other than to $\gamma\gamma$ and $\dipi\piz$, 
normalizing with branching fractions determined elsewhere~\cite{etabr}.
We subtract background from continuum by counting the yield observed
in our continuum data, scaled by luminosity and the $1/s$ dependence of the
cross-sections. The yields are listed in Table~\ref{tab:results}.

We determine the detection efficiency for all modes we study
using signal Monte Carlo (MC) samples generated with 
{\sc Evtgen}~\cite{evtgen}, including photon production
in the decay of the $\jpsi$ (``decay radiation'')~\cite{photos},
and a {\sc Geant}-based~\cite{geant} detector simulation. 
The $\pi\pi\jpsi$ samples were produced using the 
{\sc EvtGen} model {\sc VVPIPI}, with a slight
reweighting of the dipion mass distribution to 
better represent the measured spectrum~\cite{xjpsi_3m}. 
Allowing for a relative $D$-wave
component in the dipion system of the strength
found in Ref.~\cite{bes_dipi}
changes the efficiency by at most 0.1\% (relative).
The angular
distributions for the $\chicJ$ decays were generated according to
the formalism presented in Ref.~\cite{chicJMC}. In addition, 
for the $(\gamma\gamma)_{\chicJ}$ modes, we simulated the
$\chicJ$ line shape as Breit-Wigner distributions out to
$20 / 228 / 19$ ($J=0/1/2$) times the full width, 
using masses and widths from Ref.~\cite{pdg}, 
and scaled by $\egammalow^3$ as appropriate
for E1~transitions~\cite{e3weighting}. 
We compute the efficiency for $\jany$ from the weighted
sum of the individual signal MC samples, where the weight
is the relative occurrence as measured in our data. 
We will show later that the sum of our exclusive channels describes
the population observed in the inclusive selection adequately.

The following contributions to the uncertainty were evaluated: 
MC statistics (at the level of 1\%), 
tracking (0.3\% per track, added linearly), 
photon detection (0.4\% per photon, added linearly), 
MC modeling (about 1\%, additionally 
2\% for the two-photon recoil restriction for the $\chicJ$ modes) 
trigger ($0.1$-$0.4\%$, depending on decay mode), 
uncertainty of the background subtraction (stemming from all
statistical uncertainties involved). 

\begin{figure}[t]
\includegraphics*[width=6.5in]{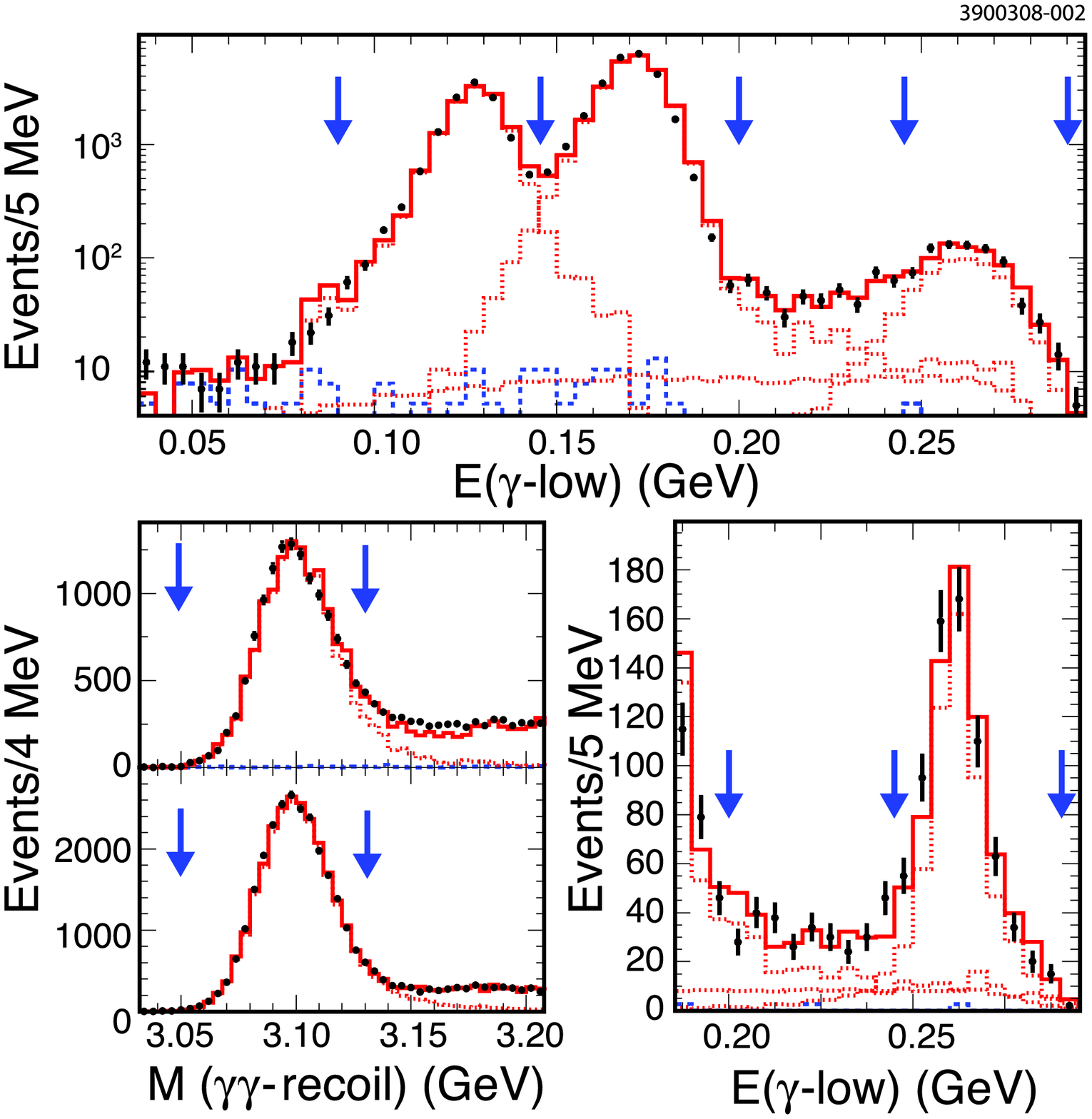}
\caption{For the decays
$\pp\to\gamma\gamma\jpsi$, 
$\egammalow$ (top) before applying the full-event
kinematic fit to the area above $\egammalow = 200\mev$,
the di-photon recoil mass for $\chi_{c2,1}$
(middle [lower] left for $J=2$ [$J=1$]), and $\egammalow$
after the full-event kinematic fit (lower right).
\symbols\ \label{fig:chicJ} }
\end{figure}

The $\chi_{c0}$ background merits some
discussion. The signal rate is heavily affected by the understanding
of the area between the $\chi_{c1}$ and $\chi_{c0}$, since whatever populates
this region also feeds into the $\chi_{c0}$ window. 
We noted in our 2005 analysis~\cite{xjpsi_3m} an excess of 
unexplained events in $\egammalow =200$-$245 \mev$ and
confirm a similar production rate here with more 
data (Fig.~\ref{fig:chicJ}). However, our simulation
of this region has improved in several ways
that significantly raise the MC expectation
for $\egammalow=200$-$245\mev$, in particular through
the aforementioned $\egammalow^3$ weighting and through
the continuation of the Breit-Wigner distribution from 
the $\chi_{c1}$ across this region
and extending into that of the $\chi_{c0}$.
The $\chi_{c1,2}$ measurements
are not affected by these considerations due to their much larger rate
(the $\chi_{c2,1,0}$ raw yields compare roughly as 20:40:1).
The events with $\egammalow=200$-$245\mev$ exhibit the 
expected behavior for $\gamma\gamma\jpsi$ events and are qualitatively
comparable to those in neighboring regions in $\egammalow$, namely: 
the quality of the $\jpsi$ fit is comparable, 
there is no indication of missing energy or momentum, 
and the lateral shower profile for photon candidates is consistent
with this assumption.

We have studied several ways to explain the data in the intermediate region 
as well as the adjacent areas: 
variation of the $\chi_{c1}$ width 
from 0.92~MeV to 1.00~MeV,
removal of the $E_\gamma^3$ weighting,
allowance for a phase-space-like component where
the $\psi(2S)$ de-excitation takes place through two non-resonant 
photons (denoted by $\ggnr$), and combinations of these.
Most of the scenarios can give a satisfactory description with
different component weightings, 
but result in substantial variations of the $\chi_{c0}$
rate due to the different apportionment of the observed yields to
signal and background.
Our data do not allow us to distinguish among the
possibilities explored,
which results in correspondingly large background-related 
contributions to the systematic uncertainty 
($\chi_{c0}$: 10\%,
$\chi_{c1}$: 1.0\%,
$\chi_{c2}$: 0.8\%). 
In any of these scenarios, more events in the $\chi_{c0}$
region are attributed to background than in Ref.~\cite{xjpsi_3m}.
We quote the result obtained
with our nominal $\chi_{c1}$ width, 
with $E_\gamma^3$ weighting, and with a phase-space-like component $\ggnr$,
as displayed in Fig.~\ref{fig:chicJ}. 
Since any sensitivity to the possible additional  $\ggnr$
component is restricted to the region between $\chi_{c0}$ and $\chi_{c1}$
(and not unambiguous there), we cannot quantify the magnitude of such
a contribution any more precisely than to say that a branching fraction
up to $1 \times 10^{-3}$ is compatible with our data.

Since we measure ratios of yields, many systematic uncertainties
cancel, most notably those related to lepton species and $\jpsi$ fitting.
To verify this, we compare ratios measured using $\jpsi \to \diel$
with those determined using $\jpsi\to\dimu$. All ratios are close to
unity; for the seven ratios involving $\dipi\jpsi$ we compute 
$\chi^2 = 3.0$ for six degrees of freedom.

Table~\ref{tab:results} shows results for the branching fraction ratios,
after combining the two measurements from $\jpsi \to \diel$ and $\dimu$.
Adding all ratios of exclusive decays to the inclusive one, 
$\Sigma_h [ (h + \jpsi) / (\jany) ]$, leads to a sum that agrees
with unity within 0.9\% or $1\sigma$. 
This favorable comparison is an indication that the contribution
from modes not covered in this analysis
(such as $\gamma \gamma \jpsi$ through $\eta_c(2S)$,
$\dipi\piz\jpsi$, {\it etc.}) is small.

\begin{figure}[t]
\includegraphics*[width=6.5in]{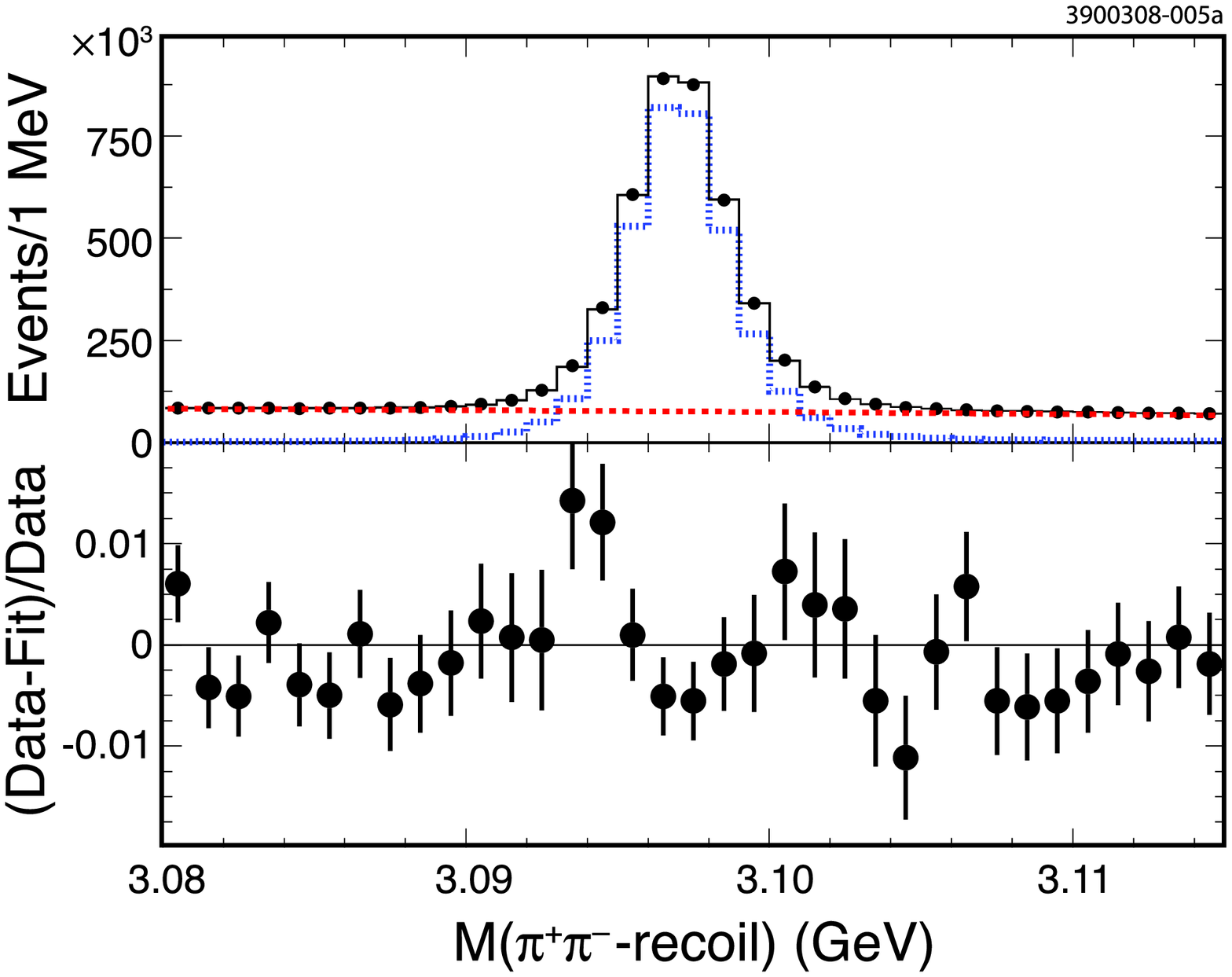}
\caption{
The dipion recoil mass spectrum for $\dipi\jpsi$, $\jpsi \to X$.
Top:
data points (black) overlaid with the fit result (solid black
curve) obtained using a (scaled) signal shape from $\dipi\jpsi$, 
$\jpsi \to \dilep$, and a second-order polynomial background shape
(red dashed curve). 
Bottom: 
the fractional difference between the fit and the data. 
\label{fig:mpipi_recoil} }
\end{figure}

We now describe a measurement of the 
$\pp \to \dipi\jpsi$, $\jpsi \to X$ production rate.
It proceeds identically to the one detailed in Ref.~\cite{cleo_jpsidilep}, 
where it was used
as the denominator in the determination of ${\cal B}(\jpsi \to \dilep)$.
The presence of the $\jpsi$ is inferred through the $\dipi$ recoil mass
spectrum. The pion candidates must consist of two oppositely charged
tracks that obey loose quality criteria, 
$p_t>150\mev$, $|\cos\theta|<0.83$,
and $m(\dipi) > 300\mev$.
The recoil mass spectrum, shown in Fig.~\ref{fig:mpipi_recoil},
is fit with a signal shape using the $m(\dipi)$-recoil spectrum
from $\dipi(\dilep)$ data events, selected with the same criteria
for the pions as the inclusive $\jpsi \to X$ sample 
(overlaid in Fig.~\ref{fig:mpipi_recoil}), and a second-order polynomial
background to extract the yield of $\dipi (X)_{\jpsi}$ events: 
$3.851 \times 10^6$. The fit has a confidence level of $98.8\%$.
The efficiency determination is potentially 
hampered by the lack of knowledge of many $\jpsi$ branching
fractions.  In practice, however,
it is found to be nearly independent of
$\jpsi$ decay mode, although it can
be expected to depend weakly upon
the track multiplicity, as detailed in Ref.~\cite{cleo_jpsidilep}.
We therefore use a sample approximating the spectrum in data with a 
weighted sum of signal MC of various multiplicities, including well-measured
decays like $\jpsi\to\dilep$. 
The agreement between data and the sum of MC predictions thus obtained is 
good~\cite{cleo_jpsidilep}.
The detection
efficiency determined from the weighted sum of individual efficiencies
is found to be $40.16\%$.
Systematic uncertainties stem from
uncertainties in the efficiency weights and modeling, translating into 0.7\%,
sensitivity to the fit range in the signal yield extraction, 0.2\%.
The result for the number of $\pp \to \dipi\jpsi$
events produced is $(9.589 \pm 0.020 \pm 0.070)\times 10^6$.

The number of $\psi(2S)$ is measured in a manner similar
to that described in Ref.~\cite{number_psiprimes}.  
Hadronic event candidates are identified by requiring three tracks 
($\ntrk\geq 3$)
and restrictions on the following quantities:
summed energies from tracks and showers, $\evis$, 
amounting to at least $0.3 \ecm$, 
and furthermore for $\ntrk=3,4$ a
summed energy deposition in the calorimeter of more than $0.15\ecm$,
and either this sum less than $0.75\ecm$
or the highest individual shower less than $0.75 \ebeam $.
The $z$-component of the event
vertex, $z_{\mathrm{vtx}}$, must be within $5\cm$ of the beam spot center.

The continuum contamination consists of
$\diel \to \ell^+\ell^-$ ($\ell\equiv e,\ \mu$, or $\tau$),
light hadron production through $\diel$ annihilation
or two-photon collisions, radiative returns
to the $J/\psi$, and $J/\psi$ decays from the
extended tail of the $J/\psi$ Breit-Wigner distribution. 
It is subtracted statistically by scaling
the yield of events passing the selection
criteria from the off-resonance data sample.
The scale factor between off- and on-resonance yields is the ratio
of luminosities (measured using the process 
$\diel \to \gamma\gamma$~\cite{DHadForLumi})
multiplied by a $1/s$ dependence for the cross-section
behavior.
We subtract non-collision events statistically using an extrapolation 
from the tail of the $z_{\mathrm{vtx}}$ into the 
signal region. 
Other effects are negligible.

\begin{figure}[t]
\includegraphics*[width=6.5in]{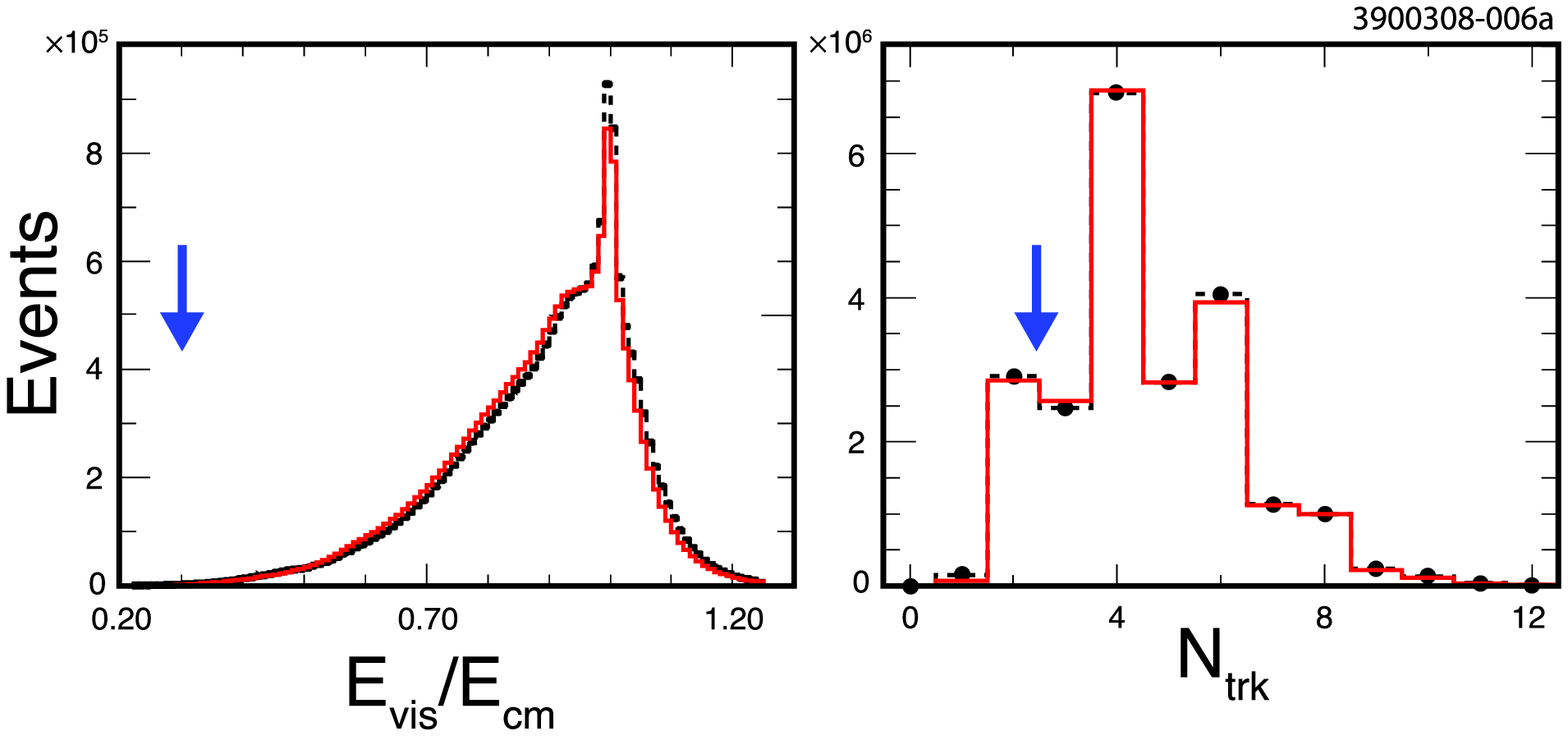}
\caption{Scaled energy in the event $\evis$
(left)
and number of tracks (right), for data (dashed black line / points) 
and simulation (solid red line). The right distribution has
been obtained with the selection based on $\ntrk \geq 1$. 
\label{fig:npsiprime} }
\end{figure}

In a MC simulation we determine the
detection efficiency for $\psi(2S)$ decay events to be 76\%.
The underlying MC generator settings incorporate current branching fraction
determinations~\cite{pdg} for $\psi(2S)$, $\chicJ$, and $\jpsi$ decays, 
and for the remainder employs JETSET~\cite{jetset}.
The agreement between data and MC simulation
can be judged from Fig.~\ref{fig:npsiprime}.

To evaluate the accuracy of the estimated
efficiency and background subtractions,
we explore three different
scenarios where we vary the requirements on
track multiplicity ($\geq 1$ to $\geq 4$) with appropriate 
background suppression criteria modifying the aforementioned energy balances, 
with detection efficiencies ranging between 91\% and 65\%, respectively. 
We find a variation of 2\% relative to the nominal setting, which we take as
a systematic uncertainty. The largest contributors are the dependence
on the trigger requirements (1.8\%), followed by the tracking and
energy settings (0.9\%).

The procedure, when applied to a portion of the earlier-taken CLEO-c
$\psi(2S)$ data, results in a slight reduction in the result compared to
the method described in Ref.~\cite{number_psiprimes}. 
This is understood to be due to updated settings of the MC generator
(thereby modifying the detection efficiency), 
a change in the continuum background subtraction
(using CLEO-c continuum data), and other improvements in the detector
description. 

The number of $\psi(2S)$ decays thus determined is
$(27.36 \pm 0.57)\times 10^6$, with a relative
systematic uncertainty of $2\%$. The statistical uncertainty is
negligible.
This leads to a branching fraction ${\cal B}(\pp \to \dipi\jpsi) = 
(35.04 \pm 0.07 \pm 0.77)\%$. This result is more precise than the one
presented in Ref.~\cite{xjpsi_3m} and also somewhat higher. This is chiefly
due to an improved background and signal efficiency treatment in the inclusive
$\psi(2S)$ count.

The last column of Table~\ref{tab:results} shows the absolute 
branching fractions, obtained by multiplying the entries in
the penultimate column by $\bpm$
and appropriate cancellation of correlated
systematic uncertainties. One independent set of numbers from
the data displayed in the table is $\bpm$ together with the
ratios to $\bpm$.
Some derived
quantities may be computed, with the correlations properly taken
into account. 
We find ${\cal B}(\piz\jpsi) / {\cal B}(\eta\jpsi) =
(3.88 \pm  0.23   \pm  0.05)\%$.
Using 
${\cal B}(\psi(2S) \to \gamma \chi_{cJ}) = (9.3 \pm 0.4)\%$, 
$(8.8 \pm 0.4)\%$, 
and
$(8.1 \pm 0.4)\%$~\cite{pdg} for $J=0,1,2$, 
we obtain 
${\cal B}( \chi_{cJ} \to \gamma\jpsi ) = $
$( 1.35\pm 0.07 \pm 0.14 \pm 0.06)\%$,
$(40.5\pm 0.3 \pm 1.4 \pm 1.8)\%$,
$(24.1\pm 0.2 \pm 0.9 \pm 1.2)\%$, respectively,
where the first uncertainty is statistical, the second systematic
from this analysis, and the third from the input branching fraction.
Our result for ${\cal B}(\chi_{c0} \to \gamma\jpsi)$ is considerably
smaller than found in Ref.~\cite{xjpsi_3m} and consistent
with other determinations~\cite{pdg}. The reduction in this rate is
a direct consequence of improved treatment of $\chi_{c0}$ backgrounds. 

With ${\cal B}(\psi(2S) \to \gamma \chicJ, \gamma \eta_c)$ and 
${\cal B}(\psi(2S) \to \dilep)$ ($\ell = e, \mu, \tau$), our
results imply~\cite{pdg} 
${\cal B}(\psi(2S) \to$ light hadrons$) = (15.4 \pm 1.5)\%$,
or 2.9~standard deviations higher than an extrapolation arrived
at by scaling ${\cal B}(\jpsi \to$ light hadrons$)$ by
the ratio of leptonic branching ratios, $(12.45 \pm 0.35)\%$.
Here, all $\psi(2S)$ results are taken
from the branching fraction fit values in Ref.~\cite{pdg}. 

In summary, we have studied the exclusive decays 
$\pp \to \jpsi + h$ ($h = \pi^+\pi^-$, $\pi^0\pi^0$,
$\eta$, $\pi^0$) and $\pp \to \gamma\chi_{cJ}\to\gamma\gamma\jpsi$
transitions, $\jpsi \to \diel$ and $\dimu$,
with a similar strategy applied to all channels.
The analysis is complemented by a study of the inclusive mode 
$\pp \to \jany$. 
We have determined branching ratios between exclusive
modes on the one hand and between exclusive modes and
$(\jany)$ on the other. 
We have also measured the branching fraction
$\pp \to \dipi\jpsi$ 
using the dipion recoil mass spectrum, which facilitates
transformation of 
the ratios relative to $\dipi\jpsi$ into absolute branching
fractions. Further quantities are derived.
The precision of all quantities
given here improves upon previous measurements.
The results presented here supersede those from Ref.~\cite{xjpsi_3m}.

We gratefully acknowledge the effort of the CESR staff 
in providing us with excellent luminosity and running conditions. 
This work was supported by 
the A.P.~Sloan Foundation, 
the National Science Foundation, 
the U.S. Department of Energy, 
the Natural Sciences and Engineering Research Council of Canada, and 
the U.K. Science and Technology Facilities Council.

\clearpage

\begin{table*}
\caption{For each channel:
the number of events observed in $\jpsi \to \dimu$ after background
subtraction and 
the detection efficiency ratio $r_h^\mu \equiv
\epsilon( \psi(2S) \to h + \jpsi^{\dimu} ) /
\epsilon( \psi(2S) \to \jany^{\dimu} ) $; 
the same for $\jpsi \to \diel$;
the ratio of branching fractions ${\cal B}(\pp \to h+\jpsi$ and
${\cal B}(\pp \to \jany)$;
the same with respect to $\bpm$; absolute branching fractions.
\label{tab:results}
}
\footnotesize
\setlength{\tabcolsep}{0.45pc}
\begin{tabular}{cccccccc} 
\hline
\hline
\rule[-1.5mm]{0mm}{4.5mm}
 Channels  
    & $N^\mu$ & $r_h^\mu$
    & $N^e$ & $r_h^e$
    & ${\cal B}/{\cal B}_{\mathrm{any}}$ (\%)
    & ${\cal B}/\bpm $ (\%)
    & ${\cal B}$ (\%) 
    \\ 
\hline 
$\pi^+\pi^-\jpsi $  
 & 302030 & 0.80
 & 263372 & 1.01
 &$ 56.04 \pm  0.09   \pm  0.62 $
 & $\equiv 100 $
 &$ 35.04 \pm 0.07 \pm 0.77 $
 \\ 
$\pi^0\pi^0\jpsi $  
 & 32249 & 0.17
 & 28746 & 0.22
 &$ 28.29 \pm  0.12   \pm  0.56 $
 &$ 50.47 \pm  0.22   \pm  1.02 $
 &$ 17.69 \pm  0.08     \pm   0.53$
 \\ 
$\eta \jpsi $  
 & 9819 & 0.27
 & 8590 & 0.33
 &$ 5.49 \pm  0.06   \pm  0.09 $
 &$ 9.79 \pm  0.10   \pm  0.15 $
 &$ 3.43   \pm  0.04     \pm   0.09$
 \\ 
$\pi^0 \jpsi $  
 & 289 & 0.19
 & 238 & 0.25
 &$ 0.213 \pm  0.012   \pm  0.003 $
 &$ 0.380 \pm  0.022   \pm  0.005 $
 &$ 0.133   \pm  0.008     \pm  0.003 $
 \\ 
$\gamma(\gamma\jpsi)_{\chi_{c0}} $ 
 & 308 & 0.22
 & 253 & 0.28
 &$ 0.201 \pm  0.011   \pm  0.021 $
 &$ 0.358 \pm  0.020   \pm  0.037 $
 &$ 0.125   \pm  0.007     \pm  0.013 $
 \\ 
$\gamma(\gamma\jpsi)_{\chi_{c1}} $  
 & 13244 & 0.34
 & 11619 & 0.44
 &$ 5.70 \pm  0.04   \pm  0.15 $
 &$ 10.17 \pm  0.07   \pm  0.27 $
 &$ 3.56  \pm  0.03    \pm  0.12 $
 \\ 
$\gamma(\gamma\jpsi)_{\chi_{c2}} $  
 & 6616 & 0.31
 & 5768 & 0.40
 &$ 3.12 \pm  0.03   \pm  0.09 $
 &$ 5.56 \pm  0.05   \pm  0.16 $
 &$ 1.95  \pm  0.02    \pm  0.07 $
 \\ 
$\jany$ 
 & 676889 & $\equiv 1 $
 & 466153 & $\equiv 1 $
 & $\equiv 100$
 &$ 178.4 \pm  0.3   \pm  2.0 $ 
 &$62.54   \pm  0.16     \pm  1.55$ \\
\hline
\hline
\end{tabular} 
\end{table*}

\end{document}